\documentclass[twocolumn]{aastex61}
\pdfoutput=1 
\usepackage{amsmath,amstext}
\usepackage[T1]{fontenc}
\usepackage{apjfonts} 
\usepackage[figure,figure*]{hypcap}

\shorttitle{AASTeX 6.1 Template}
\shortauthors{M. M. Woods et al.}

\begin{document}

\title{The triggering of the 29-March-2014 filament eruption}

\author{Magnus M. Woods}
\affiliation{Mullard Space Science Laboratory, University College London, Dorking, Surrey, RH5 6NT, UK}

\author{Satoshi Inoue}
\affiliation{Institute for Space-Earth Environmental Research\,(ISEE)/Nagoya University Furo-cho, Nagoya, Aichi 464-8601, Japan}

\author{Louise K. Harra}
\affiliation{Mullard Space Science Laboratory, University College London, Dorking, Surrey, RH5 6NT, UK}

\author{Sarah A. Matthews}
\affiliation{Mullard Space Science Laboratory, University College London, Dorking, Surrey, RH5 6NT, UK}

\author{Kanya Kusano}
\affiliation{Institute for Space-Earth Environmental Research\,(ISEE)/Nagoya University Furo-cho, Nagoya, Aichi 464-8601, Japan}

\author{Nadine M. E. Kalmoni}
\affiliation{Mullard Space Science Laboratory, University College London, Dorking, Surrey, RH5 6NT, UK}

\begin{abstract}
The X1 flare and associated filament eruption occurring in NOAA Active Region\,12017 on SOL2014-03-29 has been the source of intense study. In this work, we analyse the results of a series of non linear force free field extrapolations of the pre and post flare period of the flare. In combination with observational data provided by the IRIS, Hinode and SDO missions, we have confirmed the existence of two flux ropes present within the active region prior to flaring. Of these two flux ropes, we find that intriguingly only one erupts during the X1 flare. We propose that the reason for this is due to tether cutting reconnection allowing one of the flux ropes to rise to a torus unstable region prior to flaring, thus allowing it to erupt during the subsequent flare. 
\end{abstract}


\section{Introduction}\label{intro}
The rapid releases of magnetic energy observed as solar flares have long been associated with the eruption of plasma from the solar atmosphere. Prior to flaring and eruption, the materials that subsequently erupt can be observed as structures known as filaments. The plasma composing these filaments is thought to be suspended in magnetic structures known as flux ropes \citep[e.g.][]{Priest1989, vanB&Martens1989}.
The eruptions of these filaments are commonly thought to be driven by either ideal instabilities; such as kink instability \citep{torokkliem2005} or torus instability \citep{bateman1978, kliemtorok2006}, or by reconnection driven processes such as magnetic breakout \citep{antiocos1999} and tether cutting reconnection \citep{moorelabonte1980, moore2001}. Recent work by \cite{ishihguro2017} investigates the double arc instability\,(DAI). This instability, which is controlled by the current flowing in the flux rope, produced as a result of tether cutting reconnection, and not by decay index as in the case of the torus instability. This reliance on internal magnetic structure rather than the external field can allow a flux rope to rise at a lower height than torus unstable case. This is significant as this increase in height driven by the DAI can allow the lower altitude flux rope to rise and erupt if it then enters a torus unstable region. In the actual solar atmosphere it is likely that these processes act upon flux ropes at varying stages prior to and during eruption. \cite{inoue2016} for example presents a detailed scenario in which a flare triggering process can lead to tether cutting reconnection, which can then in turn deliver the flux rope into a torus unstable region where it then erupts.

On 29 March 2014, active region\,(AR)\, 12017 produced an X1 flare, with an associated filament eruption. This event provided unprecedented simultaneous observations of all stages of the flare from numerous observatories, providing coverage across all layers of the solar atmosphere. This has made this flare a source of intense study \citep[e.g.][]{Judge2014, matthews2015, li2015, Battaglia2015, young2015, Liu2015_mgii, aschwanden2015, KH2016, rubiodecosta2016}. \cite{Kleint2015} used observations taken by  the \textit{Interferometric BIdimensional Spectrometer}\, (IBIS) to investigate the eruption of the filament. This work revealed the presence of  consistent blue shifts of $2\,-\,4\,km\,s^{-1}$ along the filament in the hour prior to flaring. Additionally these observations also indicate the presence of two filaments within the AR\,12017, only one of which erupts during the the X1 flare. 
\cite{Woods2017} investigated the pre-flare period of this flare in detail, through the use of the Hinode/EIS and IRIS spectrometers, revealing strong transient blue shifts along the filament 40 mins before the onset of flaring. This work also utilised non-linear force free magnetic field (NLFFF) extrapolations to determine the presence of a magnetic flux rope associated with the filament present in AR 12017, focusing on the evolution over the preceding 24 hours.

The aim of this current work is investigate the triggering of the flare and subsequent filament eruption seen in AR\,12017 to complete the understanding of the pre-flare period of this iconic flare. To this end, we have produced a series of NLFFF extrapolations, focusing on the time period directly prior to and after the X1 flare in-order to investigate the origin of magnetic structure of the flux rope and the possible triggers for its destabilisation and subsequent eruption.

\section{Observations and Method}\label{obs}

The analysis presented in this paper utilises the data from several satellite observations of the 29 March 2014 X1 flare. The GOES soft X-ray lightcurve for this event is shown in Figure 1. The \textit{Extreme Ultraviolet Imaging Spectrometer}\, \citep[EIS;][]{culhane07} on board the \textit{Hinode}\,\citep[][]{kosugi07} spacecraft, was observing the AR for several hours prior to flaring. The observing program used for these observations utilised the 2" slit and raster steps of 4" to produce a field of view\,(FOV) of 42"\,x\,120". In this work we analyse the coronal Fe\,{\textsc{xii}} 195\,\AA\ emission line and pseudo-chromospheric He\,{\textsc{ii}} 256\,\AA\ . These data were fitted with single Gaussian profiles (using the solarsoft procedure, \textit{eis\_auto\_fit}), with rest wavelengths being determined experimentally, due to the lack of absolute wavelength calibration in the data. This was done by selecting a small region of quiet sun in each raster, fitting this and assuming the mean centroid velocity to be the rest velocity of the line. Doppler velocities and non-thermal velocities\,(V$_{nt}$) were calculated in the manner described in \cite{Woods2017}. 

\textit{Hinode}'s \textit{Solar Optical Telescope}\,\citep[SOT;][]{Tsuneta2008} was also observing AR 12017 in the hours prior to the X-flare. The SOT Filtergram\,(FG) was operating in shutterless mode between 14:00:31\,UT and 18:18:50\,UT. Ca \,\textsc{ii} H images were recorded with a cadence of 33\,secs, and a FOV of 55''\,x\,55''. Na\,\textsc{i} images were captured with a 16\,sec cadence and an FOV of 30''\,x\,81''. These data were aligned to the first image in the sequence in order to correct for spacecraft-jitter, and were then subsequently differentially rotated to 17:00\,UT and aligned to the 17:00\,UT Helioseismic Magnetic Imager line-of-sight magnetogram, to maintain mutual spatial alignment with the other data sources. We also utilised the SOT Spectropolarimeter\,(SP) scan of the AR produced between 17:00 and 17:55\,UT. 
 
 The \textit{Helioseismic and Magnetic Imager} \citep[HMI;][]{scherrer12} on board the \textit{Solar Dynamics Observatory}\,\citep[SDO;][]{pensell12} provides the observations of the photospheric magnetic field utilised in this paper. Vector magnetograms prepared in the Spaceweather HMI Active Region Patch (SHARP) format \citep{bobra2014}, were used to calculate non-linear force free field\,(NLFFF) extrapolations using the magnetohydrodynamic relaxation method presented in \cite{inoue2014} and \cite{inoue2016}. This method seeks to find suitable force-free fields that satisfy the lower boundary conditions, set by the photospheric magnetic fields.  We first extrapolate the potential field only from the $B_z$ component on the photosphere, which is uniquely determined \citep{sakurai1982}. Next, we gradually change the horizontal magnetic fields ($B_{{\rm pot}, x}, B_{{\rm pot}, y}$) on the lower boundary, which are potential components extrapolated from $B_z$, to match the observed horizontal fields, ($B_{{\rm obs}, x}, B_{{\rm obs}, y}$). During this process on the bottom boundary while the magnetic fields are fixed with the potential field at other boundaries, we solve following equations inside of a numerical box until the solution converges to a quasi-static state, 

   \begin{equation}
   \rho=|{\bf B}|
   \label{den_eq}
   \end{equation}

    \begin{equation}
    \frac{\partial {\bf v}}{\partial t} 
                         = - ({\bf v}\cdot{\bf \nabla}){\bf v}
                           + \frac{1}{\rho} {\bf J}\times{\bf B}
                              + \nu{\bf \nabla}^{2}{\bf v},
   \label{eq_of_mo}    
   \end{equation}

  \begin{equation}
  \frac{\partial {\bf B}}{\partial t} 
                        =  {\bf \nabla}\times({\bf v}\times{\bf B}
                        -  \eta{\bf J})
                        -  {\bf \nabla}\phi, 
  \label{in_eq}
  \end{equation}

  \begin{equation}
   {\bf J} = {\bf \nabla}\times{\bf B},
  \label{Am_low}
  \end{equation}
  
  \begin{equation}
  \frac{\partial \phi}{\partial t} + c^2_{\rm h}{\bf \nabla}\cdot{\bf B} 
    = -\frac{c^2_{\rm h}}{c^2_{\rm p}}\phi,
  \label{div_eq}
  \end{equation}
  
where $\rho$ is pseudo plasma density, ${\bf B}$ the magnetic flux density, ${\bf v}$ the velocity, ${\bf J}$ the electric current density, and $\phi$ the convenient potential to reduce errors derived from ${\bf \nabla}\cdot {\bf B}$ \citep{dedner2002}, respectively. $\nu$ is a viscosity term fixed at $1.0 \times 10^{-3}$, and the coefficients $c_h^2$, $c_p^2$ in Equation (\ref{div_eq}) are also fixed with constant values, 0.04 and 0.1, respectively.  The resistivity is given as $\eta = \eta_0 + \eta_1 |{\bf J}\times{\bf B}||{\bf v}|/{\bf |B|}$ where $\eta_0 = 5.0\times 10^{-5}$ and $\eta_1=1.0\times 10^{-3}$ in  non-dimensional units. The second term is introduced to accelerate the relaxation to the force-free field particularly in weak field region. Further details of the NLFFF extrapolation method are described in \cite{inoue2014} and \cite{inoue2016}. 
In the extrapolations presented in this work the numerical box covers an area of $230.4 \times 168.75 \times 230.4(\rm Mm^3)$ which is given as $1.0 \times 0.78125 \times 1.0$ in non-dimensional units. The region is divided into $320 \times 250 \times 320$ grids which is result of $2 \times 2$ binning process of the original SHARPS vector magnetic field in the photosphere.
   
HMI line-of-sight\,(LOS) magnetograms as well as images from SDOs \textit{Atmospheric Imaging Assembly}\,\citep[AIA;][]{lemen12} are also used to provide context images for the observations in various wavelengths, as well as providing a suitable reference for co-alignment between the different instruments. 

\section{Results}\label{results}
\begin{figure*}[]
\centerline{\includegraphics[width=1.0\textwidth,clip=]{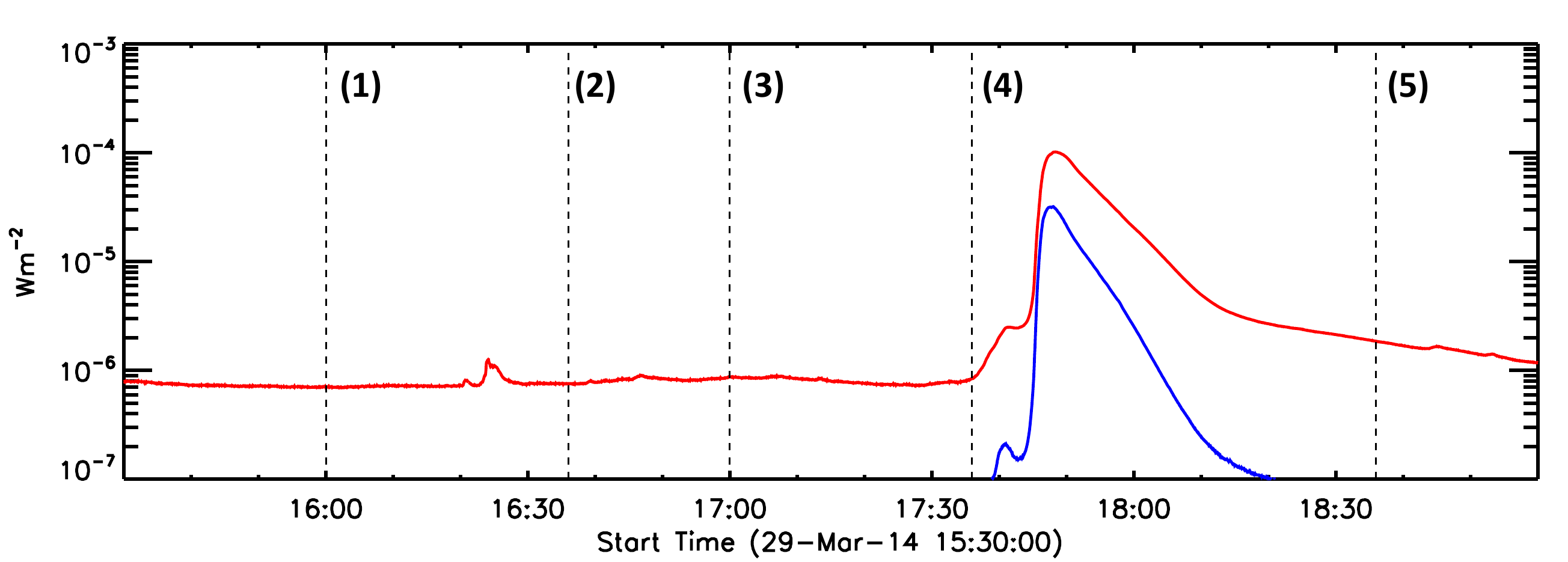}}
\caption{GOES light curve of the soft X-ray flux from 29 March 2014 15:30\,UT. The X1 flare peaked at17:48\,UT. The times of the 5 NLFFF extrapolations are marked. Extrapolations 1\,-\,3 examine the pre-flare magnetic environment, 4 details the configuration at 15:36\,UT the time of flare initiation, and the final extrapolation 5 shows the post flare magnetic field. }
\label{fig: goes}
\end{figure*}

Figure\,\ref{fig: goes} shows the GOES lightcurve between 15:30\,UT and 19:00\,UT, with the times of the five NLFFF extrapolations indicated. Extrapolations 1\,-\,3 detail the evolution of the pre-flare magnetic field, extrapolation 4 shows the field configuration at the time of flare onset\,(17:36\,UT) while extrapolation 5 shows the post flare magnetic field configuration at 18:36\,UT.  
\begin{figure*}[]
\centerline{\includegraphics[width=1.0\textwidth,clip=]{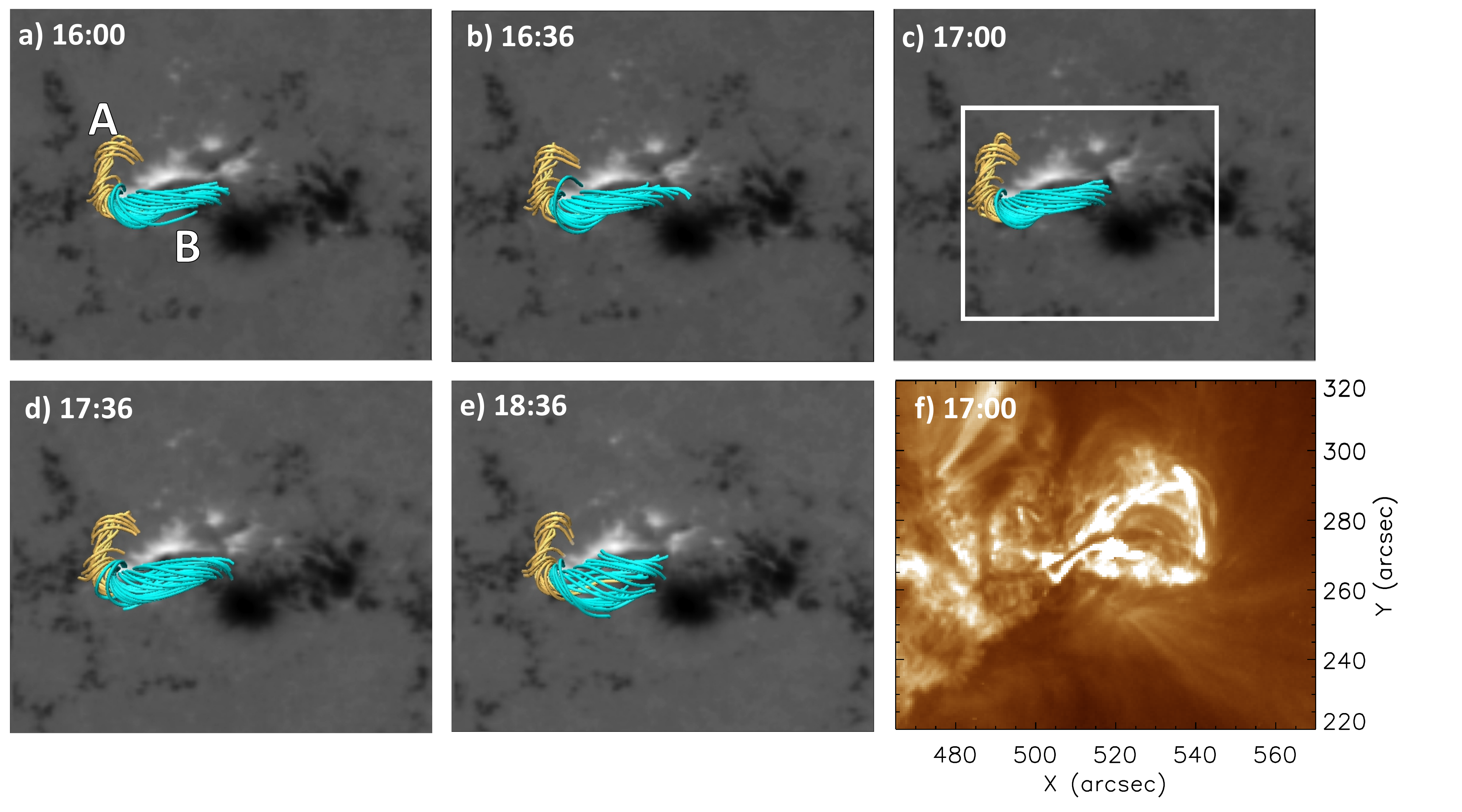}}
\caption{Here we see the five extrapolations of AR12017. Extrapolations a\,-\,c examine the pre-flare magnetic environment, d details the configuration at 15:36\,UT the time of flare initiation, and the final extrapolation e shows the post flare magnetic field. There appear to be two flux ropes within the extrapolation, the eastward structure\,(flux rope A, gold field lines) and the westward\,(flux rope B, blue field lines). There appears to be little change in the two flux ropes prior to flaring (panels a - d). However post flaring this westward feature becomes much less sheared and returns to a more potential structure in the post flare case seen in extrapolation e. Panel f shows us the AIA 192\,\AA\ image for the same field of view. The white box in panel c highlights the field of view of Figures\,\ref{fig: connectivity} and \ref{fig: twist}.}
\label{fig: extrapolations}
\end{figure*}
Figure\,\ref{fig: extrapolations} shows the results of these extrapolations in the vicinity of the polarity inversion line\,(PIL) of the active region. Figures\,\ref{fig: extrapolations} a - e chart the evolution of the flux rope from 16:00\,UT to 18:36\,UT respectively. The field lines shown are visualised within the VAPOR software \citep{clyne2005prototype,clyne2007interactive}. The field lines shown are plotted within the vicinity of the polarity inversion line, with the regions in which the field lines are plotted being kept constant in each extrapolation shown. In Figure\,\ref{fig: extrapolations} a , we see that there is a clear magnetic structure present in the active region from 16:00\,UT where field lines seem to form two separate sub structures. The eastern portion\,(A) substructure, marked by the gold field lines, appears to be more twisted, whilst the western substructure, blue\,(B), is less so. The first four extrapolations of the pre-flare period show little obvious change in the nature of these two magnetic structures. However between Figures\,\ref{fig: extrapolations} d and e we see a marked difference in the structures. We see that substructure A has maintained its twisted nature, while in contrast to this, structure B has lost its sheared nature and has become a more potential field configuration.

\begin{figure*}[]
\centerline{\includegraphics[width=1.0\textwidth,clip=]{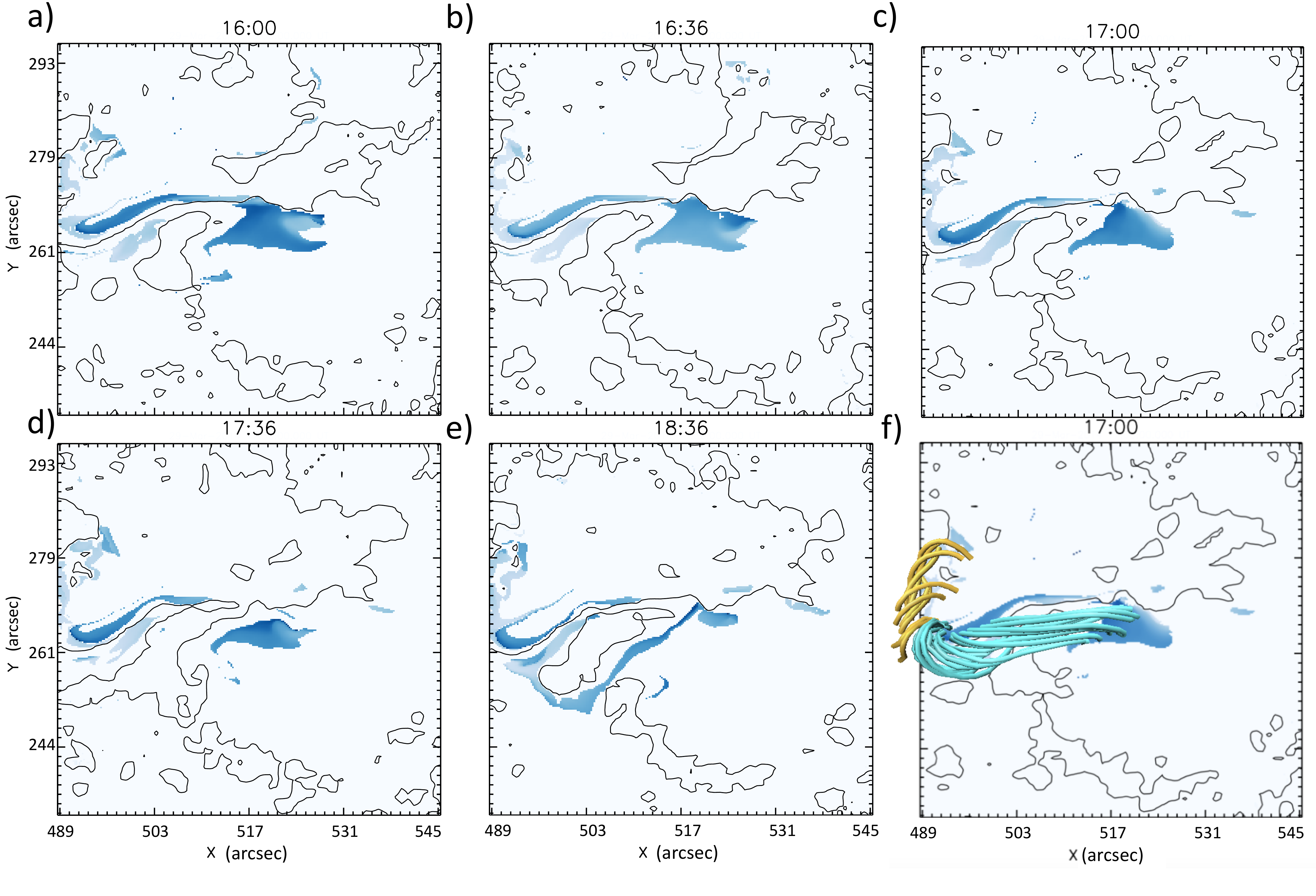}}
\caption{ This figure shows map of magnetic field line connectivity for each of the extrapolations produced. Regions which show the same colour represent the respective foot points of magnetic filed lines. We can clearly see that there are indeed two distinct systems of magnetic connectivity, on in pale blue corresponding to flux rope A, and dark blue corresponding to flux rope B. The results of these connectivity maps confirms the presence two flux ropes,  we had previously inferred from Figure\,\ref{fig: extrapolations}. Panel f shows the 17:00\,UT connectivity map with the extrapolated magnetic field lines over plotted, further confirming our conclusions. The PIL of the HMI Sharps $B_{z}$ component are also over plotted as the solid black lines. These images are plotted in CEA degrees.}
\label{fig: connectivity}
\end{figure*}
 Figure\,\ref{fig: extrapolations} f shows that the filament present in the AR prior to flaring has a strong correlation in position to the structures produced in the extrapolations. Due to the twisted nature of these features, we interpret these structures as magnetic flux ropes.
Is there any observational support that backs up the interpretation of there being two flux ropes? From the AIA images (e.g. Figure\,\ref{fig: extrapolations} f) we can only identify one filament. Whilst this is not necessarily incompatible with the findings of the extrapolation it would make it more likely that only one flux rope was present. However, the study into the 29-March-2014 X1 flare and filament eruption conducted by \cite{Kleint2015} found evidence for two separate filaments in the Ca\,\textsc{ii} 8542\,\AA\ observations made by the IBIS instrument. From these observations  \citep[Figure 2, ][]{Kleint2015} and the cartoon these authors produced of the active region and filament positions \citep[Figure 10,][]{Kleint2015} we can clearly see the resemblance to the flux ropes that are reconstructed in our extrapolations (see Figure\,\ref{fig: extrapolations}). Hence, we conclude that our extrapolation are consistent with the presence of two flux ropes, each supporting a filament within AR\,12017 prior to the X1 flare. It is important to note here, that the likely reason for our inability to observe two filaments in AIA data, is that AIA uses a broadband filter whilst the IBIS observations seen in \citet{Kleint2015} are spectrally pure.  

As further confirmation of this we mapped the connectivity of field lines within the extrapolation. To do this we calculate $\delta$, which is the distance between the footpoints of an individual field line. This distance is calculated by tracing a field line from the extrapolation to its footpoints $(x_{0}, y_{0}), (x_{1}, y_{1})$ respectively, and combined to find:
\begin{equation}
\delta = \sqrt[]{(x_{1} - x_{0})^{2} + (y_{1} - y_{0})^2}
\end{equation}
In Figure\,\ref{fig: connectivity}, we show the connectivity maps for field lines with Twist, $Tw > 0.5$. Twist is defined as
\begin{equation}
T_{\rm w}=\frac{1}{4\pi} \int \frac{{\bf \nabla}\times{\bf B}\cdot{\bf B}}{|{\bf B}|^2}dl
\end{equation}, where dl is a line element of a field line.  The colour table of the resultant plots is used to highlight regions that are connected by the same field line e.g. regions with the same colour are linked by magnetic field lines. 

The PILs of the HMI Sharps $B_{z}$ component that the extrapolations are produced from are over plotted to provide some positional information. The results of this process are shown in Figure\,\ref{fig: connectivity}. Here we can clearly see in Figures\,\ref{fig: connectivity} a - d, which chart the evolution prior to flare occurrence, that there are in fact two separate systems of magnetic field lines present. Figure\,\ref{fig: connectivity} e shows the clear changes that have occurred in the AR as a result of flaring. We can see that, as in Figure\,\ref{fig: extrapolations} e, the where we once saw flux rope B we now see a more potential magnetic structure. This can be interpreted as flux rope B erupting during the flare, and the potential field lines seen in the final extrapolation being those associated with the post flare loops. Figure\,\ref{fig: connectivity} f shows the 17:00\,UT connectivity map, with the extrapolated field lines comprising the two flux ropes. We can clearly see that the positions of the two flux ropes conform with the conclusions of the corresponding connectivity map. \\

\begin{figure*}[]
\centerline{\includegraphics[width=1.0\textwidth,clip=]{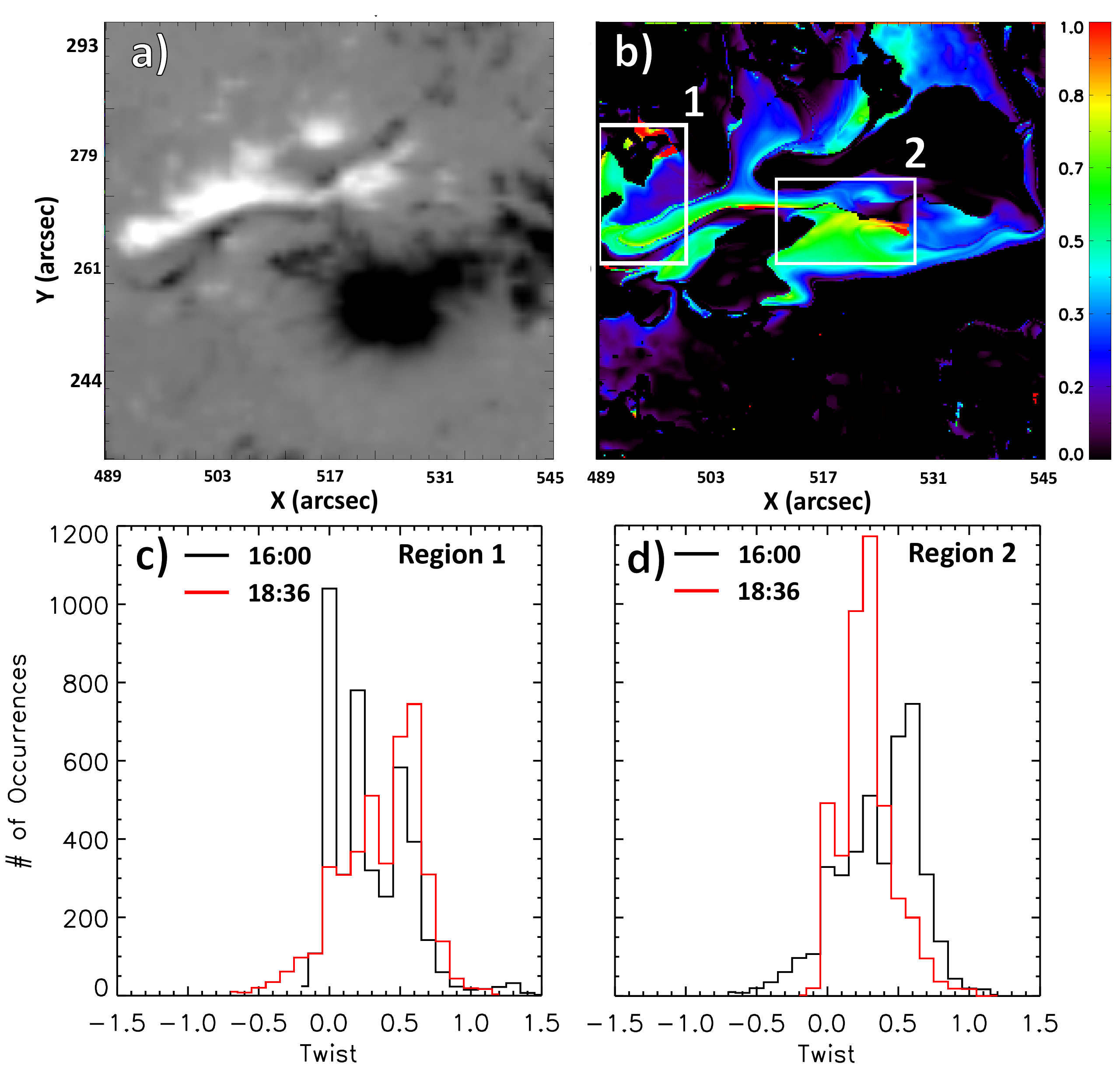}}
\caption{ Panel a shows the pre-flare HMI LOS magnetic field of the active region at 16:00\,UT. Panel b shows the twist calculated from the extrapolation at 16:00\,UT. The colour table shows twist values between 0 and 1. Panels c and d show histograms of twist in the boxes labelled 1 and 2 respectively in panel b. In both c and d the black line corresponds to twist values from the 16:00\,UT extrapolation, and red from 18:36\,UT.}

\label{fig: twist}
\end{figure*}

To quantify the changes in the magnetic field structure, we show in Figure\,\ref{fig: twist} the change in twist for two regions beneath the respective flux ropes. Panel a shows the B$_{z}$ component of the photospheric magnetic field at 16:00\,UT to act as a comparison to the twist map shown in panel b. The twist distribution at 16:00\,UT is shown in panel b, with the colour table displaying values between 0 and 1. Also highlighted are two subregions each situated beneath one of the flux ropes. The twist in each pixel of regions 1 and 2 was investigated  for the pre-flare (16:00\,UT) extrapolation and the post-flare (18:36\,UT) extrapolation. Panels c and d correspond to regions 1 and 2 respectively, with twist values corresponding to 16:00\,UT shown in black and 18:36\,UT shown in red.  What can be seen is that in the case of region 1 twist in flux rope A increases after the flare has occurred, while twist in flux rope B has decreased.   

The difference between these two separate flux ropes presents an intriguing problem: Why does flux rope B erupt, despite it being less twisted than flux rope A? 

\begin{figure*}[]
\centerline{\includegraphics[width=1.0\textwidth,clip=]{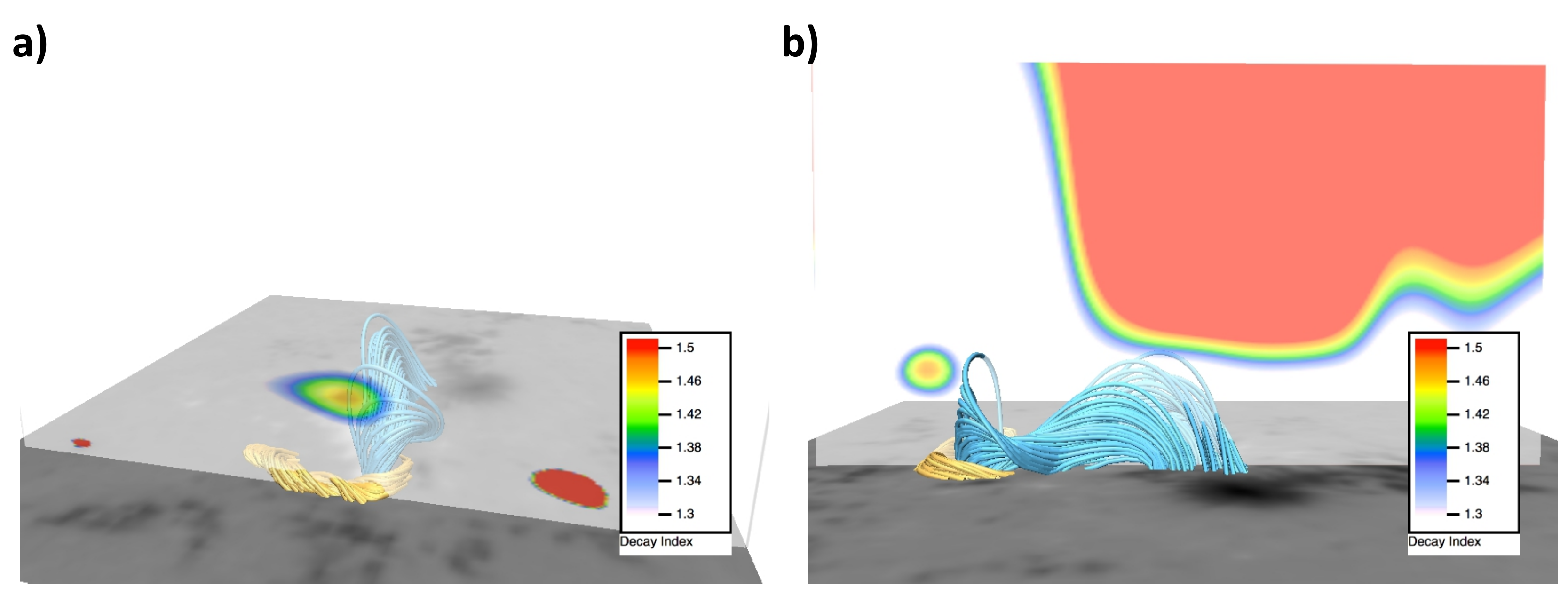}}
\caption{The spacial distribution of decay index, n, is shown over plotted on the extrapolations. We see in panel a that flux rope A, gold field lines, is very low altitude and is situated far below the torus unstable region. Panel b shows that flux rope B, blue field lines, lies at a higher altitude and thus, closer to the torus unstable region. }
\label{fig: decay}
\end{figure*}
\begin{figure*}[]
\centerline{\includegraphics[width=1.0\textwidth,clip=]{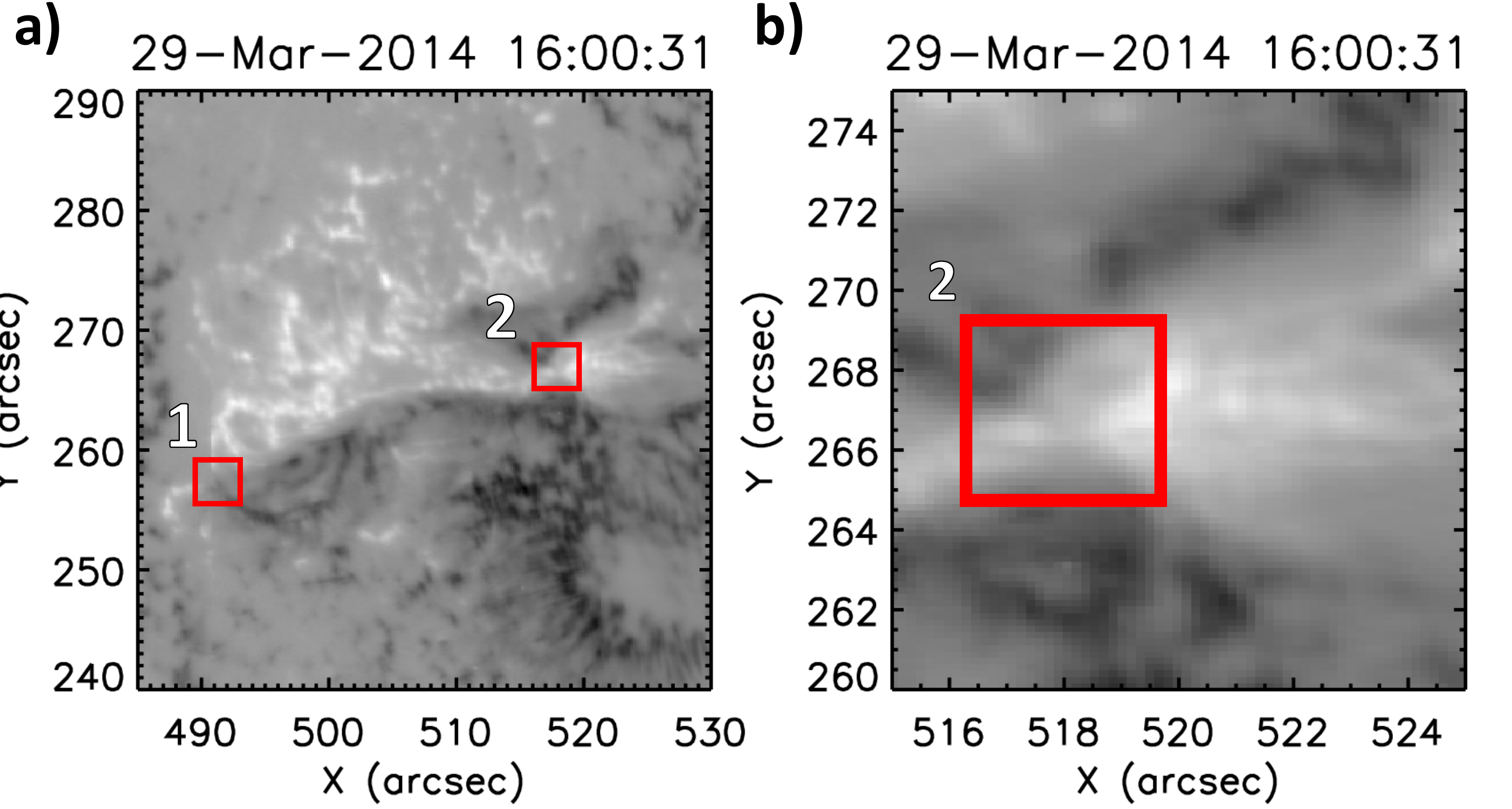}}
\caption{Stokes V map of AR\,12017. Panel a shows the full field of view, with the two sub regions of study marked by the boxes labelled 1 and 2 respectively. Panel b shows an insert of the region around region 2. This area undergoes flux cancellation. An animated version of this figure is available online, and is referred to in the text as supplementary movie 1. This animation shows the evolution of the Stokes V fields of view between 16:00\,UT and 17:35\,UT the time of flare onset.}
\label{fig: stokes_v}
\end{figure*}
\begin{figure*}[]
\centerline{\includegraphics[width=1.0 \textwidth,clip=]{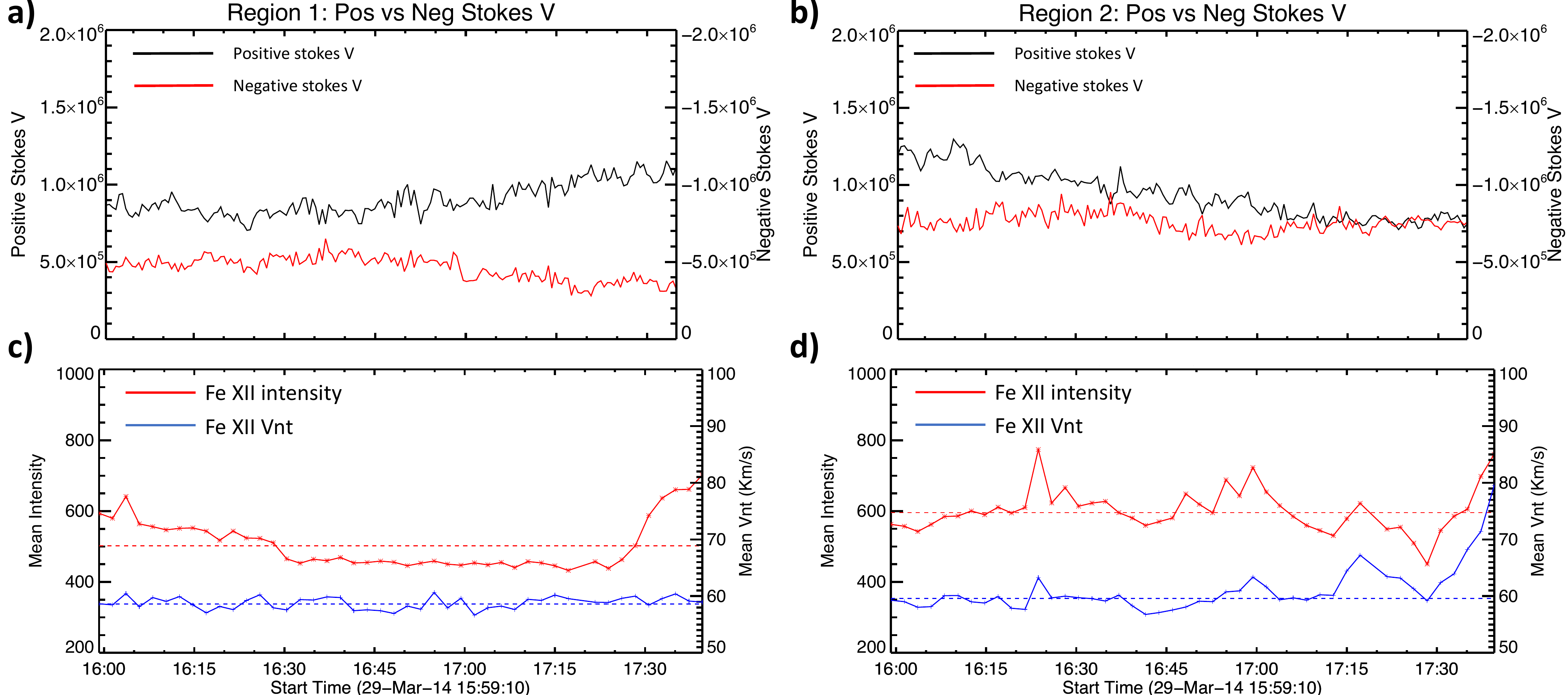}}
\caption{Panels a and b detail the evolution of positive (black) and negative (red) Stokes V for the two regions seen in Figure\,\ref{fig: stokes_v} a respectively. Evidence of flux cancellation is clearly seen in panel a between 16:30 and 17:00\,UT. Panels c and d show for the same time period the evolution of intensity (red) and non-thermal velocity (blue). We see that for region 2 (panel d), which experiences flux cancellation, there are intensity enhancements during this time as well as a trend towards increasing non-thermal velocity. This is not seen in region 1, panel c.}
\label{fig: stokes_fe}
\end{figure*}
\begin{figure*}[]
\centerline{\includegraphics[width=1.0\textwidth,clip=]{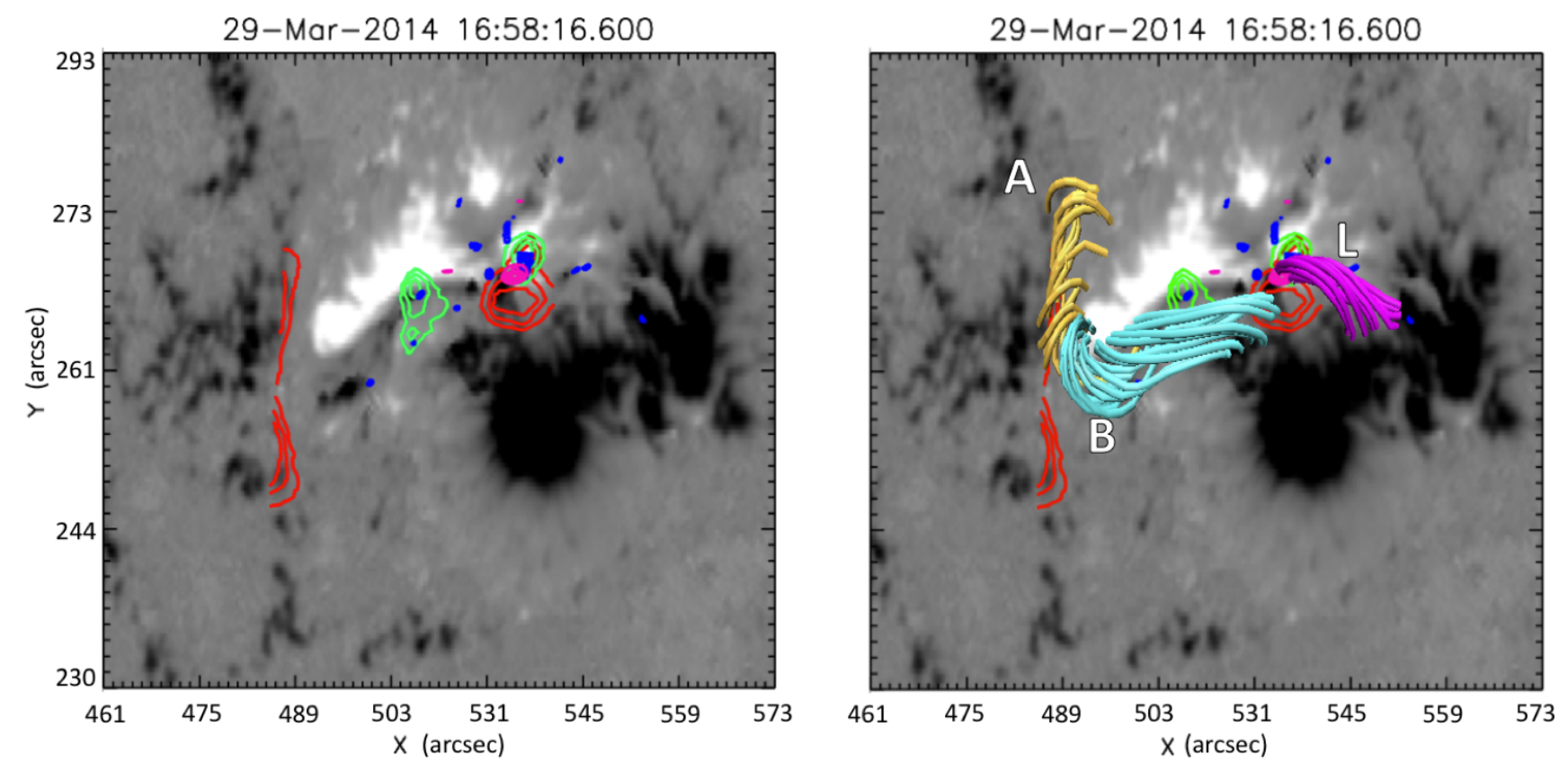}}
\caption{ Panel a shows the locations of the most intense emission in several spectral lines covering the full solar atmosphere: coronal Fe\,\textsc{xii}\,(red) and pseudo-chromospheric He\,\textsc{ii}\,(green) as observed by Hinode EIS; the transition region line Si\,\textsc{iv}\,(pink), as seen by IRIS and the chromospheric Ca\,\textsc{ii}\,(blue) observed by Hinode SOT. These data are then overlayed onto the HMI Sharps maps used in the preparation of the extrapolations. We see that most brightenings are centred upon the region where we observed the cancellation of flux in the Stokes V data.
In panel b, we see flux rope B in relation to a group of field lines (L, shown in purple). We propose that it is the interaction between flux rope B and these field lines that leads to the flux cancellation and observed brightenings. Also shown are the locations of the brightenings from panel a.}
\label{fig: overlays}
\end{figure*}

Alongside the twist of these structures, we must also consider their stability to torus instability \citep{bateman1978, kliemtorok2006}. To investigate this, the decay index, a dimensionless parameter that quantifies the gradient of magnetic field strength with height, is calculated from the extrapolation results. Decay index is given by:
\begin{equation}
    n = -\frac{\partial\ln B}{\partial\ln Z}
\end{equation}
where, B is the magnetic field strength and Z is the radius of the torus, which is equivalent to height above the photosphere. In a region where $n\geq 1.5$ the flux rope will be susceptible to torus instability. This work utilised the horizontal component of the magnetic field B in the calculation of the decay index. 
Figure \ref{fig: decay} shows the spatial distribution of the decay index calculated from the 17:00\,UT extrapolation. We can see that flux rope A (Figure\,\ref{fig: decay} panel a) is well below the region in which it would be susceptible to torus instability. Flux rope B however is much closer to a region of high decay index, see Figure\,\ref{fig: decay} panel b. The higher altitude of flux rope B and its proximity to the torus unstable region provides strong evidence as to why it was more likely to erupt during the X1 flare. However, a mechanism to propel flux rope B into the torus unstable region is still necessary. \\
To investigate the possible cause of this we turn to observational sources to see if an explanation presents itself. In Figure\,\ref{fig: stokes_v}\,(and accompanying supplementary movie 1) we see (in panel a) the Stokes V component in AR\,12017 as observed by Hinode SOT's filtergram. Supplementary movie 1 shows the evolution of the Stokes V component, which we shall use as a proxy for magnetic field throughout this work,  between 16:00\,UT and flare onset at 17:35\,UT. Two regions of interest were selected, as marked by the white boxes in Figure\,\ref{fig: stokes_v}. Each of these is located on the PIL beneath the location of the eastward\,(box 1) and westward\,(box 2) magnetic flux ropes respectively. Figures\,\ref{fig: stokes_fe} a and b show the evolution of positive and negative Stokes V within these regions. 
We can clearly see that the evolution of Stokes V in these two regions is very dissimilar. The eastward region 1 shows between 16:00\,UT and 16:35\,UT a small decrease in positive Stokes V, and an equally small increase in negative Stokes V. After this time positive and negative Stokes V mirror each other closely with positive Stokes V increasing and negative Stokes V decreasing. In contrast, the westward region 2 (Figure\,\ref{fig: stokes_fe} b) we see that positive Stokes V decreases throughout the period of observation. Between 16:00\,UT and $\sim$16:35\,UT negative Stokes V varies, in fact showing a slight increase during this time. From 16:35\,UT however, negative Stokes V is seen to drop significantly until $\sim$17:00\,UT at which point it is seen to rise once more. After this point ($\sim$17:15\,UT) the values of Stokes V stabilise and remain fairly constant until flare onset. \citet{harra2013} utilised observations of non-thermal velocity calculated from spectra obtained by Hinode/EIS to identify locations of pre-flare activity. \cite{Woods2017} also utilised this technique to identify signature, that they attributed to being most likely driven by tether cutting reconnection. In Figures\,\ref{fig: stokes_fe} c and d we show, for the same time period and areas, the evolution of Fe \textsc{xii} intensity and non-thermal velocity\,($V_{nt}$). From $\sim$16:40\,UT in region 2 (Figure\,\ref{fig: stokes_fe} d) there is an increase in intensity and the start of an upward trend of $V_{nt}$ that continues until flare onset. This timing coincides with the decrease in both positive and negative Stokes V seen in this region. In contrast there is little activity seen in either intensity or $V_{nt}$ from region 1\,(Figure\,\ref{fig: stokes_fe} a).  Supplementary movie 1 and Figure\,\ref{fig: stokes_v} b focus on the area around the region 2 where we observe the apparent flux cancellation in Figure\,\ref{fig: stokes_fe} b. 
To investigate cause of the $V_{nt}$ and intensity enhancements, data taken by several other satellites were used. Figure\,\ref{fig: overlays} shows the locations of the most intense emission in several spectral lines covering the full solar atmosphere: coronal Fe\,\textsc{xii} (195\,\AA) and pseudo-chromospheric He\,\textsc{ii} (256\,\AA) as observed by Hinode EIS; the transition region line Si\,\textsc{iv}, as seen by IRIS and the chromospheric Ca\,\textsc{ii}. These data are then overlayed onto the HMI Sharps maps used in the preparation of the extrapolations. We see that most brightenings are centred upon the region where we observed the apparent cancellation of flux in the Stokes V data. 

\section{Discussion} \label{discussion}
NLFFF modelling has confirmed that prior to the eruption two separate flux ropes are present within the active region. During the flare, flux rope B is seen to erupt whilst flux rope A does not. This is also supported by the results of the extrapolations (See Section\,\ref{results}, Figure\,\ref{fig: extrapolations}), where we see that post flaring flux rope B has been replaced with a more potential magnetic field configuration, whilst flux rope A has gained twist.  This result is somewhat surprising, as flux rope A is seen (in the pre-flare extrapolations) to have consistently higher twist that its western counterpart. One might expect that the flux rope with the highest twist would be most likely to erupt through several possible mechanisms or instabilities (e.g. kink instability etc.). So, why then in this case do we see the flux rope with lesser twist erupting counter to expectations? The answer most likely comes from the brightenings we highlighted in Figure\,\ref{fig: overlays} panel a. Here we found brightenings throughout several layers of the solar atmosphere also accompanied by enhanced $V_{nt}$ signatures. These signatures are highly suggestive of magnetic reconnection. Additionally, the brightenings are all coincident with a location of possible flux cancellation. The presence of flux cancellation is determined in Figure\,\ref{fig: stokes_fe} b, and can be clearly seen in supplementary movie 1. We interpret these observations as indications of magnetic reconnection occurring below flux rope B, consistent with the presence of tether cutting flux cancellation in the vicinity of the neutral line \citep{moore2001}.

\begin{figure*}[]
\centerline{\includegraphics[width=1.0\textwidth,clip=]{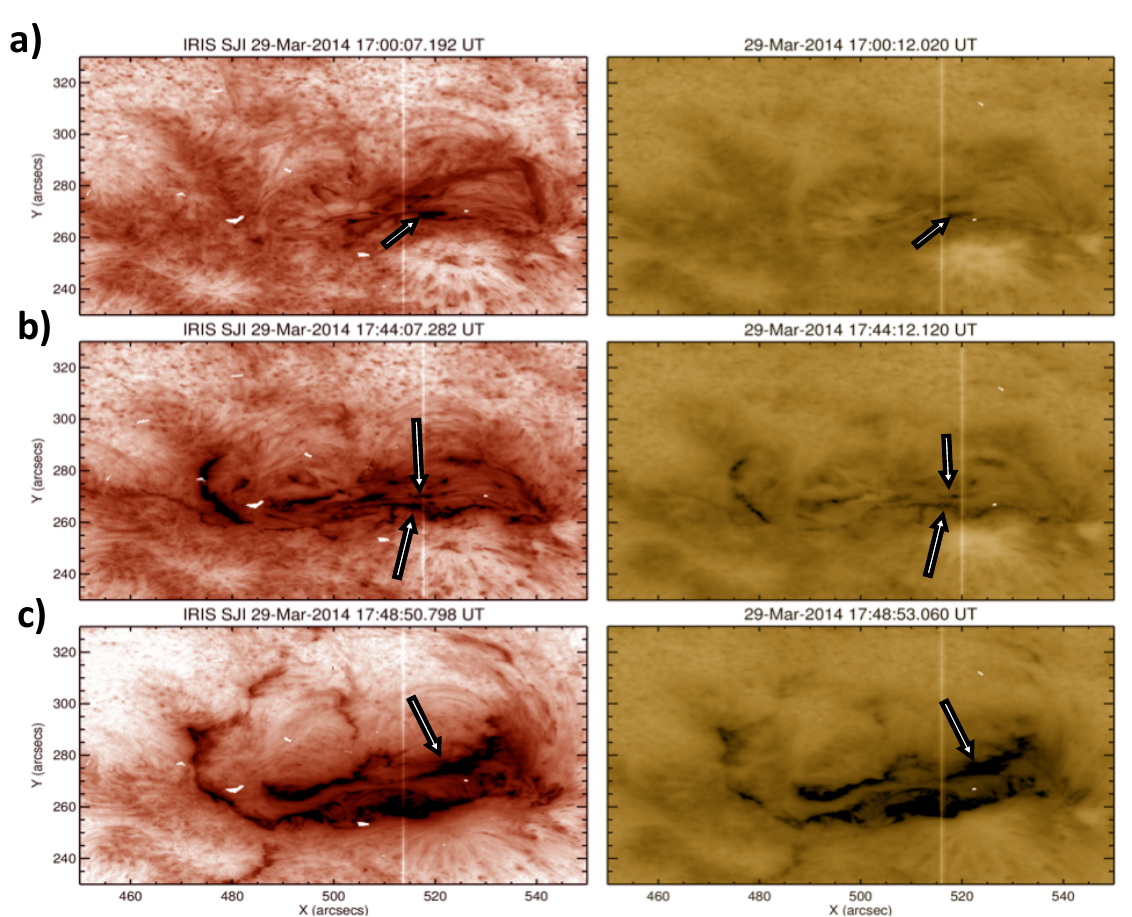}}
\caption{In this figure we see IRIS SJI images of AR\,12017. The left hand column shows 1400\,\AA\ data whilst the right shows 2796\,\AA\. In panel a we see the pre-flare brightening seen above the flux cancellation region, marked by the arrow. Panel b shows the situation during the flare at 17:45\,UT, where the arrows indicate the flare ribbons that have extended from the brightening in panel a. Panel c shows the post flare ribbons, with the arrow indicating the ribbon that resulted from the pre-flare brightening.}
\label{fig: ribbons}
\end{figure*}

Additionally, from panel b of Figure \ref{fig: overlays} we can see that the likely source of the field lines that reconnect with flux rope B leading to the flux cancellation are shown in purple. These extend from the sunspot to the region where we observe the flux cancellation beneath flux rope B. This reconnection could in turn destabilise flux rope B, eventually leading to its eruption. The absence of flux cancellation in the vicinity of flux rope A in the hour leading up to flaring could explain why it remains stable and non-eruptive in the flare itself. Further evidence for this region being heavily involved in the triggering of the X1 flare comes from examination of the flare ribbons. Figure\,\ref{fig: ribbons} shows IRIS slit jaw images\,(SJI) of the active region in 1400\,\AA\ and 2796\,\AA\ (left and right hand columns respectively). In panel a, we see the SJI data taken at 17:00\,UT. The arrow denotes the location of the bright region we discussed earlier. Panel b shows the same field of view at 17:45\,UT, during the early stages of the flare. We see that the brightening from Figure\,\ref{fig: ribbons} b has elongated slightly to form a flare ribbon and has also been  joined by a corresponding ribbon to the south. In Figure\,\ref{fig: ribbons} c, 17:48\,UT we see clearly the full flare ribbons and note (with the arrow) the position of the ribbon that resulted from the initial brightening.
As we have discussed, flux rope B is most likely destabilised by reconnection occurring in the flux cancellation region. This reconnection is most likely tether cutting reconnection, thus allowing the flux rope to rise and eventually erupt. 
But, why then does the companion flux rope A not erupt, despite its twisted nature? Firstly, from Figures\,5 and 6 we know that there is little sign of flux cancellation in the vicinity of flux rope A. Additionally we see little evidence from other sources\,(e.g. Hinode EIS, IRIS etc.) of intensity enhancements within the region of this flux rope. This clear difference to flux rope B allows us to infer that it is highly unlikely that tether cutting reconnection is occurring in flux rope A. Although this flux rope is observed to have higher twist than the flux rope B, the level of twist was found to be $\sim$1, which is below the threshold, $T_{w}\,=\,1.75$\,(\cite{torokkliem2004}), for kink instability to occur, thus giving it further stability.\\
  From the results and analysis we have presented, we propose the following scenario that leads to the triggering of the eruption of flux rope B. The interaction of flux rope B with the purple field lines shown in Figure\,\ref{fig: overlays} b leads to the onset of tether cutting reconnection between these two features. This is evidenced by the flux cancellation and brightenings seen in Figures \ref{fig: stokes_fe} and \ref{fig: overlays} respectively. This tether cutting reconnection then possibly leads to the onset of Double Arc Instability, DAI. At this point prior to flaring, flux rope B is in a region where the decay index is below the threshold necessary for Torus instability to occur. Per \cite{ishihguro2017}, current may increase in flux rope B due to the tether cutting reconnection, leading to the onset of DAI, allowing flux rope B to enter the Torus unstable regime and to erupt during the X1 flare.
\cite{Kleint2015} observed blue shifts along the filament during the slow rise flare of the X1.0 flare and during the eruption its self. The velocities observed by \cite{Kleint2015} are of order of $100s\,km\,s^{-1}$ (dependant on spectral line observed), and as such are consistent with those expected from DAI\,\citep[See;][Section 4.2 and Figure 7]{ishihguro2017}, where velocities of $320\,km\,^{-1}$ are predicted.

Flux rope A on the other hand, despite appearing to be more highly twisted than flux rope B, lacks the interaction with other magnetic fields to allow tether cutting reconnection and DAI to propel it into the torus unstable region. Thus, whilst flux rope B is able to erupt, flux rope A is non-eruptive during the X1 flare.

\section{Conclusion}\label{conclusion}
In this paper we have presented an analysis of the pre-flare period of the X1 flare that occurred in AR 12017. We produced a series of five NLFFF extrapolations in order to investigate the evolution of the magnetic field in the active region. These extrapolations not only confirmed the presence of a flux rope within the active region, but revealed that this flux rope was in fact composed of two separate flux ropes. Of these two flux ropes, only the western flux rope\,(B) erupted during the flare. Utilising observations from multiple layers of the atmosphere, in combination with Hinode SOT FG observations of the photospheric magnetic field, we discovered evidence of flux cancellation beneath the western flux rope up to one hour prior to flaring leading to reconnection. It is this reconnection that we believe destabilises the flux rope and allows its subsequent eruption during the flare.

We propose that it is tether cutting reconnection which allows flux rope B to rise slowly, possibly leading to the onset of DAI, which in turn propels the flux rope from a torus stable region to a region where it is subject to this instability. Therefore, during the X1 flare flux rope B is able to erupt from the active region. We also theorise that despite the twisted nature of the eastward flux rope\,(A), it does not erupt during the X1 flare for the following reasons: 1) the absence of destabilising flux cancellation and following tether cutting reconnection, 2) although it is twisted, the twist in the eastern flux rope is below the threshold for kink instability to occur.

\acknowledgments
The authors would like to acknowledge:
IRIS is a NASA Small Explorer mission developed and operated by LMSAL with mission operations executed at NASA Ames Research centre and major contributions to downlink communications funded by ESA and the Norwegian Space Centre. Hinode is a Japanese mission developed and launched by ISAS/JAXA, collaborating with NAOJ as a domestic partner, and NASA and STFC (UK) as international partners. Scientific operation of the Hinode mission is conducted by the Hinode science team organised at ISAS/JAXA. HMI and AIA are instruments on-board SDO, a mission for NASA's Living with a Star program. Data are provided courtesy of the SDO science team. NLFFF extrapolations were visualised using VAPOR (\cite{clyne2005prototype}, \cite{clyne2007interactive}), which is a product of the National Centre for Atmospheric Research's Computational and Information Systems Lab. We would like to that the Japan Society for the Promotion of Science\,(JSPS) for providing MMW with the JSPS summer vacation Scholarship that enabled this work. MMW also acknowledges STFC for support via their PhD Studentship. LKH, SAM and NMEK acknowledge STFC consolidated grant ST/N000722/1 for their funding. 


\bibliographystyle{yahapj}
\bibliography{references}


\end{document}